\documentclass{article}
\usepackage{spconf,amsmath,graphicx}
\usepackage{bm}
\usepackage{subcaption}
\newcommand\numberthis{\addtocounter{equation}{1}\tag{\theequation}}
\DeclareMathOperator*{\argminA}{arg\,min} 
\usepackage{mwe} 

\title{
Matthews Correlation Coefficient Loss for Deep Convolutional Networks: Application to Skin Lesion Segmentation
}
\name{Kumar Abhishek and Ghassan Hamarneh
\thanks{Corresponding author: Kumar Abhishek (kabhishe@sfu.ca)}
}
\address{School of Computing Science, Simon Fraser University, Canada\\
\texttt{\{kabhishe, hamarneh\}@sfu.ca}}
\begin{document}
%
\maketitle
\begin{abstract}
The segmentation of skin lesions is a crucial task in clinical decision support systems for the computer aided diagnosis of skin lesions. Although deep learning-based approaches have improved segmentation performance, these models are often susceptible to class imbalance in the data, particularly, the fraction of the image occupied by the background healthy skin. Despite variations of the popular Dice loss function being proposed to tackle the class imbalance problem, the Dice loss formulation does not penalize misclassifications of the background pixels. We propose a novel metric-based loss function using the Matthews correlation coefficient, a metric that has been shown to be efficient in scenarios with skewed class distributions, and use it to optimize deep segmentation models. Evaluations on three skin lesion image datasets: the ISBI ISIC 2017 Skin Lesion Segmentation Challenge dataset, the DermoFit Image Library, and the PH2 dataset, show that models trained using the proposed loss function outperform those trained using Dice loss by 11.25\%, 4.87\%, and 0.76\% respectively in the mean Jaccard index. The code is available at {https://github.com/kakumarabhishek/MCC-Loss}. 
\end{abstract}
\section{Introduction}
\label{sec:intro}

With over 5 million annual diagnoses in the USA alone~\cite{FF2020}, skin cancer is the most common form of cancer. Melanoma, the deadliest form of skin cancer representing only a small fraction of all skin cancer diagnoses, accounts for over 75\% of all skin cancer related deaths, and is estimated to be responsible for 6,850 fatalities in the USA alone during 2020 ~\cite{CancerStatistics2020}. However, studies have shown that early detection of skin cancers can lead to five-year survival rate estimates of approximately 99\%~\cite{CancerStatistics2020}, necessitating early diagnosis and treatment. Computer-aided diagnosis and clinical decision support systems for skin cancer detection are reaching human expert-levels~\cite{esteva2017dermatologist,brinker2019deep}, and a crucial step for skin lesion diagnosis is the delineation of the skin lesion boundary to separate the affected region from the healthy skin, known as lesion segmentation. The recent advances in machine and deep learning have resulted in significant improvements in automated skin lesion diagnosis, but it remains a fairly unsolved task because of complications arising from the large variety in the presentation of these lesions, primarily, shape, color, and contrast.

Medical images often suffer from the data imbalance problem, where some classes occupy larger regions in the image than others. In the case of skin lesion images, this is frequently observed when the lesion is just a small fraction of the image with healthy skin occupying the majority of the image (for example, see the first two and the last rows in Figure~\ref{fig:results}). Unless accounted for while training a deep learning-based segmentation model, such an imbalance can lead to the model converging towards a local minimum of the loss function, yielding sub-optimal segmentation results biased towards the healthy skin~\cite{salehi2017tversky}.
Cross-entropy based loss values are often a poor reflection of segmentation quality on validation sets, and therefore overlap-metric based loss functions are preferred~\cite{berman2018lovasz}.
Variations of the Dice loss~\cite{milletari2016v}, a popular overlap-based loss function modeled using the Sørensen-Dice index, have been proposed to account for class imbalance in medical image segmentation tasks~\cite{salehi2017tversky,sudre2017generalised,abraham2019novel}. Similarly, some works have proposed using a combination of a distance-based loss (e.g. cross-entropy loss) and an overlap-based loss (e.g., Dice loss) to address the data imbalance issue~\cite{wong20183d,taghanaki2019combo}. For a detailed survey on segmentation loss functions, we direct the interested readers to Taghanaki et al.~\cite{taghanaki2020deep}. The Dice loss, however, does not include a penalty for misclassifying the false negative pixels~\cite{zhang2020kappa}, affecting the accuracy of background segmentation. We therefore propose a novel loss function based on the Matthews correlation coefficient (MCC)~\cite{matthews1975comparison}, a metric indicating the correlation between predicted and ground truth labels. MCC is an informative metric even when dealing with skewed distributions~\cite{maqc2010microarray} and has been shown to be an optimal metric when designing classifiers for imbalanced classes~\cite{boughorbel2017optimal}. Motivated by these meritorious properties of MCC, in this work, we present a MCC-based loss function that operates on soft probabilistic labels obtained from a deep neural network based segmentation model, making it differentiable with respect to the predictions and the model parameters. We evaluate this loss function by training and evaluating skin lesion segmentation models on three clinical and dermoscopic skin image datasets from different sources, and compare the performance to models trained using the popular Dice loss function.

\section{Method}
\label{sec:method}

\begin{figure}[!tbp]
  \begin{subfigure}[t]{0.47\columnwidth}
    \includegraphics[width=\columnwidth]{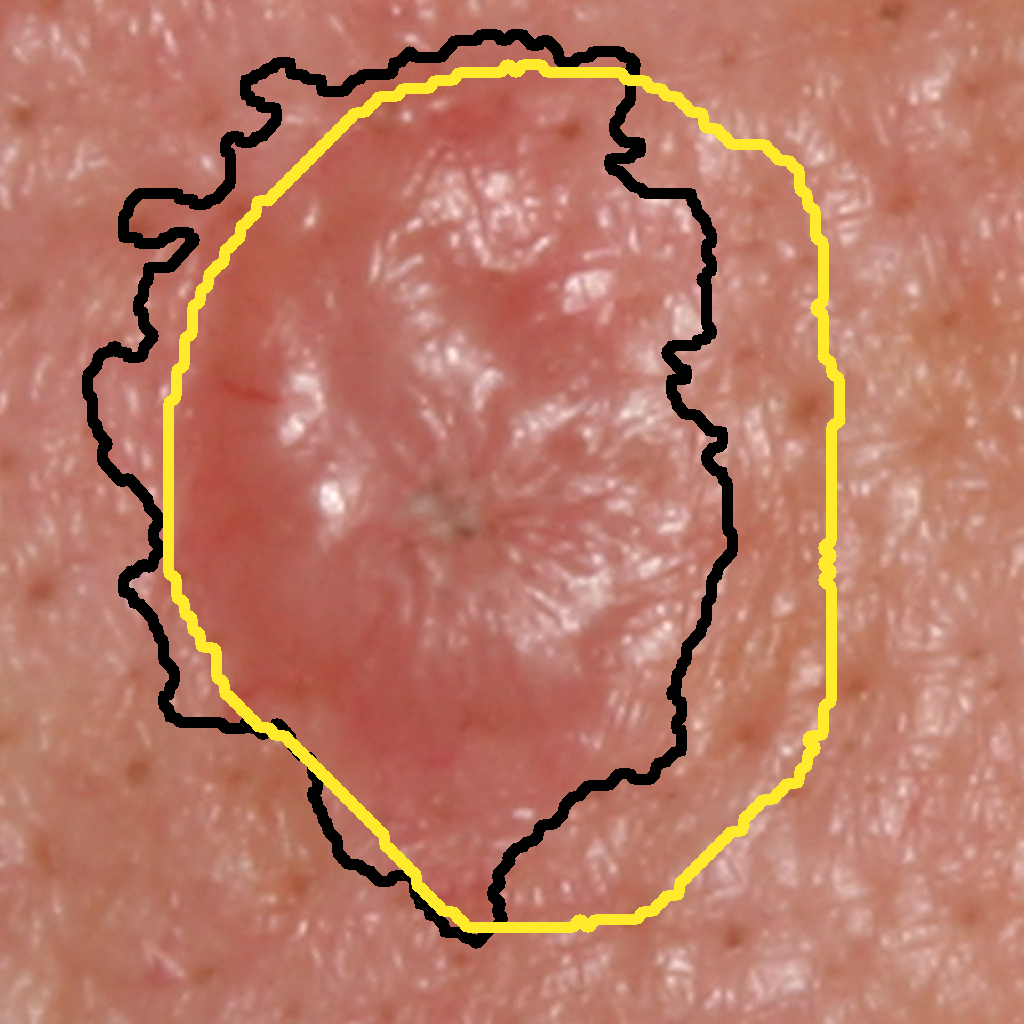}
    \caption{A skin lesion image with overlaid masks.}
  \end{subfigure}
  \hfill
  \begin{subfigure}[t]{0.47\columnwidth}
    \includegraphics[width=\columnwidth]{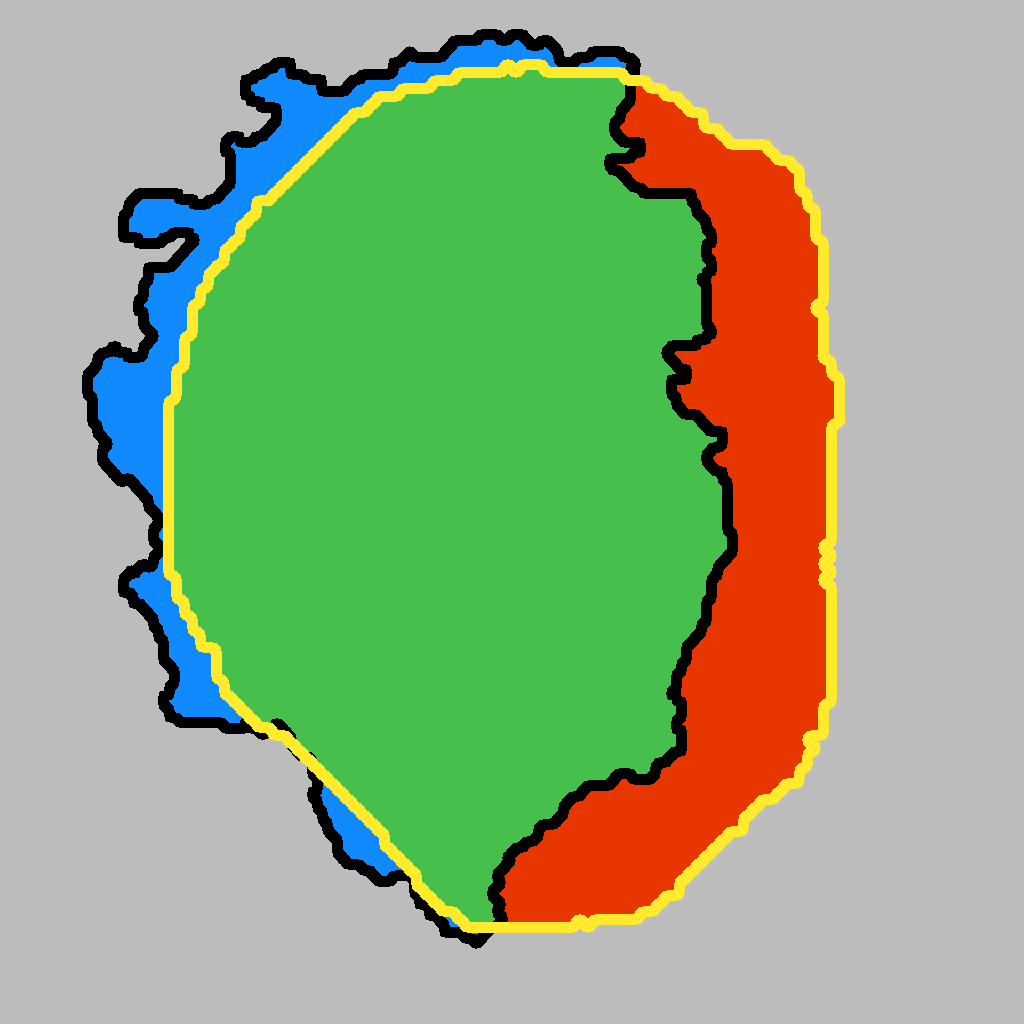}
    \caption{Overlap of the predicted and the ground truth masks.}
  \end{subfigure}
  \caption{A skin lesion image with the predicted (yellow) and ground truth (black) skin lesion segmentation masks: true positive (TP), true negative (TN), false positive (FP), and false negative (FN) predictions are denoted by green, grey, red, and blue respectively.}
  \label{fig:overlap}
\end{figure}

Consider the binary segmentation task where each pixel in an image is labeled as either foreground or background. Figure~\ref{fig:overlap} shows a skin lesion along with the corresponding ground truth and predicted segmentation masks, denoted by $Y = \{y_i\}_{i=1}^N$ and $\hat{Y} = \{\hat{y}\}_{i=1}^N$ respectively.

Consider the two popular overlap metric based loss functions: intersection-over-union (IoU, also known as Jaccard) loss and Dice loss. They are modeled using the Jaccard index and the Dice similarity coefficient (DSC), respectively, which are defined as:

\begin{equation}
    \mathrm{Jaccard} = \frac{\mathrm{TP}}{\mathrm{TP} + \mathrm{FP} + \mathrm{FN}},
\end{equation}
\begin{equation}
    \mathrm{DSC} = \frac{\mathrm{2TP}}{\mathrm{2TP} + \mathrm{FP} + \mathrm{FN}}.
\end{equation}

\noindent where true positive (TP), false positive (FP), and FN (false negative) predictions are entries from the confusion matrix. Notice that neither of these metrics penalize misclassifications of the true negative (TN) pixels, making it difficult to optimize for accurate background prediction. We instead propose a loss based on the Matthews correlation coefficient (MCC). The MCC for a pair of binary classification predictions is defined as:

\begin{equation}
\mathrm{MCC} = \frac{(\mathrm{TP} \cdot \mathrm{TN}) - (\mathrm{FP} \cdot \mathrm{FN})}{\sqrt{(\mathrm{TP}+\mathrm{FP})(\mathrm{TP}+\mathrm{FN})(\mathrm{TN}+\mathrm{FP})(\mathrm{TN}+\mathrm{FN})}}   \label{eqn:mcc_metric}
\end{equation}

MCC values range from $-1$ to $1$, with $-1$ and $1$ indicating a completely disjoint and a perfect prediction respectively. An MCC-based loss function, $\mathcal{L}_{\mathrm{MCC}}$, can be defined as: 
\begin{equation}
    \mathcal{L}_{\mathrm{MCC}} = 1 - \mathrm{MCC} \label{eqn:mcc_loss}
\end{equation}

For a differentiable loss function defined on pixelwise probabilistic predictions from the segmentation network, we define:

\begin{equation*}
    \mathrm{TP} = \sum^N_i \hat y_i y_i \ ; \ \mathrm{TN} = \sum^N_i (1-\hat y_i) (1-y_i);
\end{equation*}
\begin{equation}
    \mathrm{FP} = \sum^N_i \hat y_i (1-y_i) \ ; \ \mathrm{FN} = \sum^N_i (1-\hat y_i) y_i, \numberthis \label{eqn:conf_entries}
\end{equation}

\noindent where $\hat y_i$ and $y_i$ denote the prediction and the ground truth for the $i^{\mathrm{th}}$ pixel in the image. Dropping the summation limits for readability and plugging values from Eqn.~\ref{eqn:conf_entries} into Eqn.~\ref{eqn:mcc_metric}, we have:

\begin{equation}
    \mathcal{L}_{\mathrm{MCC}} = 1 - \frac{\sum \hat y_i y_i - \frac{\sum \hat y_i \sum y_i}{N}}{f(\hat y_i, y_i)},
\end{equation}

\begin{align}
    f(\hat y_i, y_i) = \sqrt{
    \begin{aligned}
        & \sum \hat y_i \sum y_i - \frac{\sum \hat y_i (\sum y_i)^2}{N} \\ & - \frac{(\sum \hat y_i)^2 \sum y_i)}{N} + (\frac{\sum \hat y_i \sum y_i}{N})^2.
    \end{aligned}
    }
\end{align}

The gradient of this formulation computed with respect to the $i^{\mathrm{th}}$ pixel in the predicted segmentation is:

\begin{equation}
    \frac{\partial \mathcal{L}_{\mathrm{MCC}}}{\partial \hat y_i} = \frac{1}{2} \frac{g(\hat y_i, y_i)}{\left(f(\hat y_i, y_i)\right)^{\frac{3}{2}}} - \frac{y_i - \frac{\sum y_i}{N}}{f(\hat y_i, y_i)},
\end{equation}

\begin{align}
    \begin{aligned}
        g(\hat y_i, y_i) &= \left(\sum \hat y_i y_i - \frac{\sum \hat y_i \sum y_i}{N}\right) \cdot \left(\sum y_i -\right. \\ & \left.\frac{(\sum y_i)^2}{N}\right.  \left.- 2\frac{\sum \hat y_i \sum y_i}{N} + 2\frac{\sum \hat y_i (\sum y_i)^2}{N} \right).
    \end{aligned}
\end{align}

Finally, we optimize the deep segmentation model $f(\cdot)$ using error backpropagation as:
\begin{equation}
    \Theta^* = \argminA_\Theta \mathcal{L}_{\mathrm{MCC}}\left(f(X,\Theta), Y\right),
\end{equation}
\noindent where $\hat{Y} = f(X, \Theta)$ denotes the segmentation for input image $X$ predicted by the model parameterized by $\Theta$.

\section{DATASETS AND EXPERIMENTAL DETAILS}
\label{sec:experiments}

Given that the goal of this work is to demonstrate the efficacy of using an MCC-based loss to optimize deep convolutional neural networks for segmentation, we use U-Net~\cite{ronneberger2015u} as baseline the segmentation network. The U-Net architecture consists of symmetric encoder-decoder networks with skip connections carrying features maps from corresponding layers in the encoder to the decoder, thus smoothing the loss landscape~\cite{li2018visualizing} and tackling the problem of gradient vanishing~\cite{taghanaki2020deep}.

We evaluate the efficacy of optimizing segmentation networks using the MCC-based loss on three clinical and dermoscopic skin lesion image datasets, namely the ISIC ISBI 2017 dataset, the DermoFit Image Library, and the PH2 dataset. The ISIC ISBI 2017 dataset~\cite{codella2018skin} contains skin lesion images and the corresponding lesion segmentation annotations for three diagnosis labels: benign nevi, melanoma, and seborrheic keratosis. The dataset is partitioned into training, validation, and testing splits with 2000, 150, and 600 image-mask pairs respectively. The DermoFit dataset~\cite{ballerini2013color} and the PH2 dataset~\cite{mendoncca2013ph} contain 1300 and 200 image-mask pairs belonging to 10 and 3 diagnosis classes, respectively. We randomly partition the DermoFit and the PH2 datasets into training, validation, and testing splits in the ratio of $60:10:30$.

For each dataset, we train two U-Net based segmentation models, one trained with the Dice loss ($\mathcal{L}_{\mathrm{Dice}}$) and another with the MCC loss ($\mathcal{L}_{\mathrm{MCC}}$) and compare their performance. All the images and the ground truth segmentation masks are resampled using nearest neighbor interpolation to $128 \times 128$ resolution using Python's SciPy library. All networks are trained using mini-batch stochastic gradient descent with a batch size of 40 (largest batch size that could fit in the GPU memory) and a learning rate of $1e-3$. While training, we use on-the-fly data augmentation with random horizontal and vertical flips and rotation in the range $[-45^{\circ}, 45^{\circ}]$. All models are implemented using the PyTorch framework. For evaluating the segmentation performance, we report the metrics used by the ISIC challenge, namely, pixelwise accuracy, Dice similarity coefficient, Jaccard index, sensitivity, and specificity, and use the Wilcoxon two-sided signed-rank test for statistical significance.

\section{RESULTS AND DISCUSSION}
\label{sec:results}

\begin{figure}[ht]
\centering
\includegraphics[width=0.8\columnwidth]{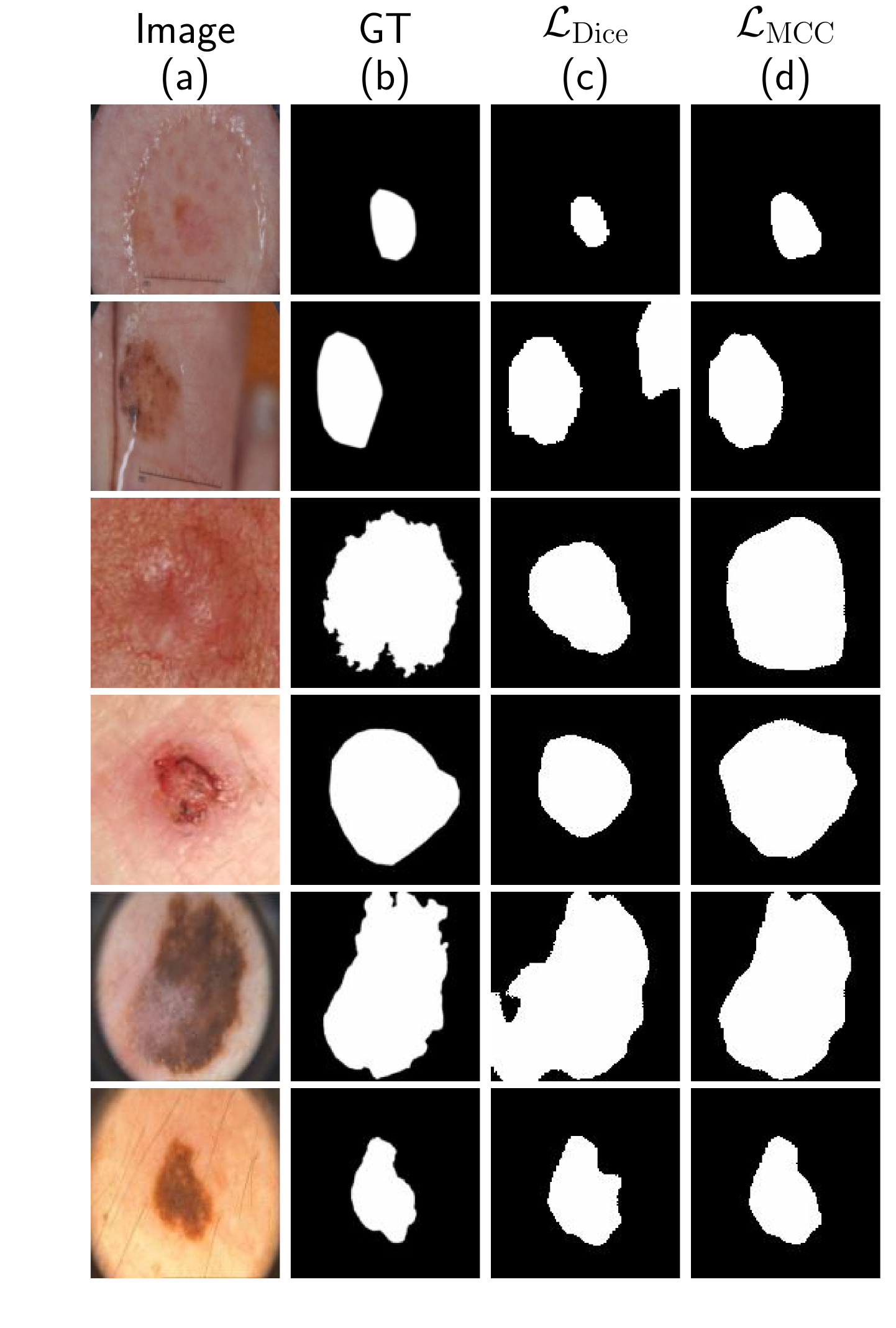}
\caption{Qualitative skin lesion segmentation results for the three datasets with (a) the original image, (b) ground truth mask, and lesion segmentation predictions with models trained with (c) $\mathcal{L}_{\mathrm{Dice}}$ and (d) $\mathcal{L}_{\mathrm{MCC}}$. The first two rows contain images from the ISIC 2017 dataset, the next two from the DermoFit dataset, and the last two from the PH2 dataset.}
\label{fig:results}
\end{figure}

\begin{table*}[ht]
\centering
\caption{Quantitative results for segmentation models trained with the two loss function evaluated on the test partitions of the ISIC 2017 (600 images), DermoFit (390 images), and PH2 (60 images) datasets (mean $\pm$ standard error). \textbf{***} and \textbf{*} denote statistical significance of the Jaccard index at $p<0.001$ and $p<0.05$ respectively.}
\label{tab:results}
\resizebox{\textwidth}{!}{%
{\renewcommand{\arraystretch}{1.2}
\begin{tabular}{c|cc|cc|cc}
\hline
\textbf{Dataset}       & \multicolumn{2}{c|}{\textbf{ISIC 2017\textsuperscript{***}}} & \multicolumn{2}{c|}{\textbf{DermoFit\textsuperscript{***}}} & \multicolumn{2}{c}{\textbf{PH2\textsuperscript{*}}}   \\ \hline
\textbf{Loss Function} & $\mathcal{L}_{\mathrm{Dice}}$          & $\mathcal{L}_{\mathrm{MCC}}$           & $\mathcal{L}_{\mathrm{Dice}}$          & $\mathcal{L}_{\mathrm{MCC}}$          & $\mathcal{L}_{\mathrm{Dice}}$        & $\mathcal{L}_{\mathrm{MCC}}$         \\ \hline
Dice          & $0.7781 \pm 0.0086$   & \bm{$0.8384 \pm 0.0070$}   & $0.8437 \pm 0.0043$   & \bm{$0.8709 \pm 0.0030$}  & $0.8888 \pm 0.0027$ & \bm{$0.8937 \pm 0.0020$} \\ 
Jaccard       & $0.6758 \pm 0.0095$   & \bm{$0.7518 \pm 0.0084$}   & $0.7418 \pm 0.0056$   & \bm{$0.7779 \pm 0.0041$}  & $0.8051 \pm 0.0038$ & \bm{$0.8112 \pm 0.0032$} \\ 
Accuracy      & $0.9029 \pm 0.0053$   & \bm{$0.9217 \pm 0.0046$}   & $0.9024 \pm 0.0063$   & \bm{$0.9137 \pm 0.0023$}  & $0.9219 \pm 0.0032$ & \bm{$0.9300 \pm 0.0022$} \\ 
Sensitivity   & $0.7470 \pm 0.0091$   & \bm{$0.8130 \pm 0.0080$}   & $0.8080 \pm 0.0063$   & \bm{$0.8799 \pm 0.0040$}  & $0.9132 \pm 0.0027$ & \bm{$0.9155 \pm 0.0029$} \\ 
Specificity   & $0.9683 \pm 0.0031$   & \bm{$0.9710 \pm 0.0029$}   & \bm{$0.9533 \pm 0.0024$}   & $0.9300 \pm 0.0030$  & $0.8852 \pm 0.0074$ & \bm{$0.9075 \pm 0.0053$} \\ \hline
\end{tabular}
}
}
\end{table*}

\begin{figure*}[!htbp]
\centering
\includegraphics[width=0.9\textwidth]{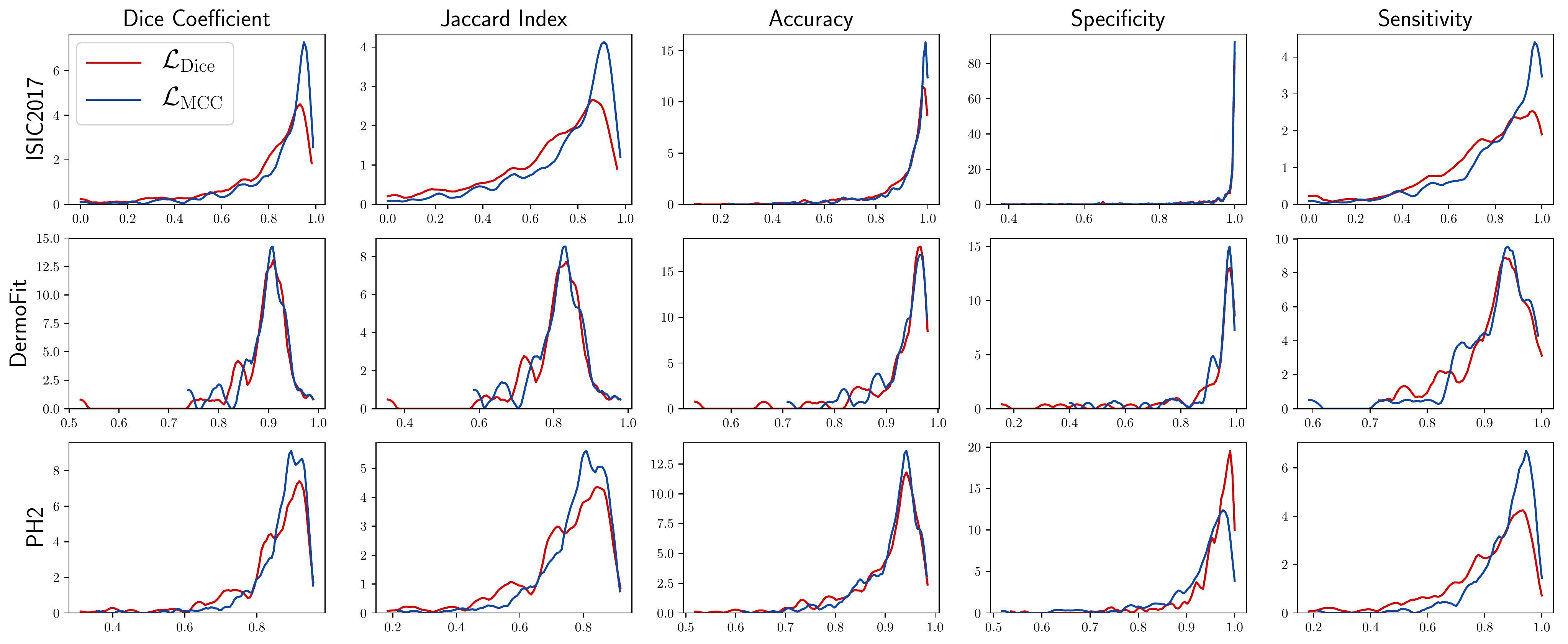}
\caption{Kernel density estimate plots for all the metrics from models trained using the Dice ($\mathcal{L}_{\mathrm{Dice}}$) and the MCC ($\mathcal{L}_{\mathrm{MCC}}$) losses evaluated on the three datasets.}
\label{fig:KDE}
\end{figure*}

To compare the performance of the models trained using the two losses, we present both qualitative and quantitative results for all the three datasets. Table~\ref{tab:results} contains the 5 evaluation metrics for two models for all the datasets. We see that models trained with $\mathcal{L}_{\mathrm{MCC}}$ outperform those trained with $\mathcal{L}_{\mathrm{Dice}}$ on all metrics for all the datasets (except the specificity on DermoFit), with improvements in both sensitivity and specificity values.
Even for the DermoFit dataset, we observe that the model trained with $\mathcal{L}_{\mathrm{MCC}}$ achieves a better trade-off between sensitivity and specificity (0.8799 and 0.9300 obtained using $\mathcal{L}_{\mathrm{MCC}}$ versus 0.8080 and 0.9533 obtained using $\mathcal{L}_{\mathrm{Dice}}$). 
The models trained with $\mathcal{L}_{\mathrm{MCC}}$ improve the mean Jaccard index by 11.25\%,  4.87\%, and 0.76\% on ISIC 2017, DermoFit, and PH2 datasets, respectively.
Additionally, the performance on the ISIC 2017 dataset is within 1\% of the Jaccard index achieved by the top 3 entries on the challenge leaderboard
even with a vanilla U-Net architecture and without using any post-processing, external data, or an ensemble of prediction models.

Next, to demonstrate the improvement in the segmentation prediction, we plot kernel density estimates of all the metrics for the three datasets in Figure~\ref{fig:KDE} using the Epanechnikov kernel to estimate the respective probability density functions. The plots have been clipped to the observed values for the corresponding metrics. We observe higher peaks (i.e., higher densities) at higher values for models trained using $\mathcal{L}_{\mathrm{MCC}}$. The improvements in Jaccard index for ISIC 2017 and DermoFit are statistically significant at $p<0.001$, and for PH2 at $p<0.05$, possibly explained by the small sample size (60 test images).

Figure~\ref{fig:results} presents 6 images sampled from the test partitions of the three datasets as well as the corresponding ground truth segmentation masks and the predicted segmentation masks using the two models. The images capture a wide variety in the appearance of the lesions, in terms of the size and the shape of the lesion, the lesion contrast with respect to the surrounding healthy skin, and the presence of artifacts such as markers and dark corners. We observe that the models trained with $\mathcal{L}_{\mathrm{MCC}}$ produces more accurate outputs, with considerably fewer false positive and false negative predictions.

\section{CONCLUSION}
\label{sec:conclusion}

We proposed a novel differentiable loss function for binary segmentation based on the Matthews correlation coefficient that, unlike IoU and Dice losses, has the desirable property of considering all the entries of a confusion matrix including true negative predictions. Evaluations on three skin lesion image datasets demonstrate the superiority of using this loss function over the Dice loss for training deep semantic segmentation models, with more accurate delineations of the lesion boundary and fewer false positive and negative predictions. Interestingly, we observed in our experiments that a model trained using Dice loss yielded an inferior Dice coefficient upon evaluation as compared to a model trained using MCC-based loss, and is similar to the observations of Zhang et al.~\cite{zhang2020kappa}, therefore warranting further investigation.
Other future directions would be generalizing this loss function for $K$ classes using entries from a $K \times K$ confusion matrix and evaluating this loss function on other medical imaging modalities.



\section{Compliance with Ethical Standards}
\label{sec:ethics}

This research study was conducted retrospectively using human subject data made available in open access by the International Skin Imaging Collaboration: Melanoma Project for the ISIC 2017 dataset~\cite{codella2018skin} and the ADDI (Automatic computer-based Diagnosis system for Dermoscopy Images) Project for the PH2 dataset~\cite{mendoncca2013ph}, and through an academic license from the University of Edinburgh for the DermoFit dataset~\cite{ballerini2013color}. Ethical approval was not required as confirmed by the respective licenses attached with the data.

\section{Acknowledgments}
\label{sec:acknowledgments}

Funding for this work was provided by the Natural Sciences and Engineering Research Council of Canada
(NSERC RGPIN-06752) and Canadian Institutes of Health Research (CIHR OQI-137993). The authors are grateful to the NVIDIA Corporation for donating Titan X GPUs used in this research.

\bibliographystyle{IEEEbib}
\bibliography{refs}

\end{document}